\documentclass[sigconf]{acmart}
\AtBeginDocument{%
  }
\copyrightyear{2026}
\acmYear{2026}
\setcopyright{cc}
\setcctype{by}
\acmConference[IH '26]{Interactive Health Conference}{July 05--08, 2026}{Porto, Portugal}
\acmBooktitle{Interactive Health Conference (IH '26), July 05--08, 2026, Porto, Portugal}
\acmDOI{10.1145/3786579.3804932}
\acmISBN{979-8-4007-2422-0/2026/07}

\begin{document}
\title{Design Guidelines for Game-Based Refresher Training of Community Health Workers in Low-Resource Contexts}

\author{Arka Majhi}
\email{arka.majhi@iitb.ac.in}
\affiliation{%
  \institution{IIT Bombay}
  \city{}
  \country{India}
}
\orcid{https://orcid.org/0000-0002-5057-1878}

\author{Aparajita Mondal}
\email{aparajita.mondal@tuni.fi}
\affiliation{%
  \institution{Tampere University}
  \city{}
  \country{Finland}
}
\orcid{https://orcid.org/0000-0003-4609-2249}

\author{Satish B. Agnihotri}
\email{sbagnihotri@iitb.ac.in}
\affiliation{%
  \institution{IIT Bombay}
  \city{}
  \country{India}
}
\orcid{https://orcid.org/0000-0002-0703-3185}

\renewcommand{\shortauthors}{ Majhi et al.}

\begin{abstract}
    Community Health Workers (CHWs) play a critical role in delivering primary healthcare services in low-resource settings, yet sustaining their training and performance remains a persistent challenge. Prior research has explored digital and game-based approaches for CHW training. However, limited work has synthesized longitudinal design insights into generalizable guidelines for interactive health interventions. Building on a four-year design-based research program involving multiple game-based refresher training systems, including quiz-based mobile apps, physical and augmented reality games, card-based games, and location-based games, we examine which design guidelines support sustained engagement, learning transfer, and contextual appropriateness in CHW training.

    We conducted a mixed-methods analysis across deployments with Accredited Social Health Activists and Anganwadi Workers in India, including interviews, field observations, and usage logs. Through thematic synthesis, we derive eight design guidelines addressing contextual realism, adaptive learning, hybrid interaction, social motivation, explainability, professional identity, and ethical considerations. Our findings contribute actionable design knowledge for researchers and practitioners developing interactive health interventions in low-resource healthcare contexts.
\end{abstract}

\ccsdesc[500]{Human-centered computing~Ubiquitous and mobile computing}
\ccsdesc[500]{Human-centered computing~Empirical studies in HCI}
\ccsdesc[500]{Applied computing~Health informatics}
\ccsdesc[500]{Human-centered computing~Interaction design}
\ccsdesc[500]{Human-centered computing~Human computer interaction (HCI)}
\ccsdesc[500]{Human-centered computing~Field studies}

\keywords{Community Health Workers, Serious Games, Interactive Health, Game-Based Training, Design Guidelines, Low-Resource Contexts, HCI4D}

\maketitle

\section{Introduction}

Researchers have explored digital \citep{HealthSpokenTutorial} and game-based approaches to health training and education. Serious games and playful learning environments offer opportunities for increased engagement, experiential learning, and reflection \citep{Graafland2014, DeRibaupierre2014, Laamarti2014, DeWit-Zuurendonk2011}. Prior studies \citep{Minzi2023} have demonstrated that game-based systems can support knowledge acquisition and motivation in healthcare education contexts \citep{sayoof2024}. However, much of this work has focused on clinicians, students, or controlled laboratory environments, with comparatively less attention to frontline health workers in community settings. Existing research has largely emphasized short-term usability and learning outcomes rather than long-term engagement, learning transfer, and contextual appropriateness.

In India, researchers have conducted design-based research investigating game-based refresher training for CHWs. This has led to multiple interactive systems, including instructional illustrations \cite{Tulaskar2020}, quiz-based mobile applications \cite{Majhi2021,Shah2017}, physical and augmented reality games \cite{Majhi2022}, digital–physical hybrid card-based games \cite{Majhi2024HCI_International, Majhi2024CHIPlay}, location-based games for malnutrition mapping and performance awareness \cite{Majhi2024GoodIT}, discussions on challenges for developing them \cite{Majhi2026IHBoF} and framing design guidelines \cite{Majhi2026IHSP}. Each system addressed specific design questions related to engagement, embodiment, social learning, and situated practice. While these studies demonstrated the feasibility of game-based training, they also revealed recurring patterns of conflicts, such as competition versus cooperation, automation versus explainability, and engagement versus ethical concerns related to monitoring and surveillance.

Despite growing interest in serious games and interactive health interventions, there remains a lack of empirically grounded design knowledge to guide the design of such systems for CHWs in low-resource contexts. Existing frameworks for game-based learning and health behavior change provide general guidance but do not sufficiently account for the socio-cultural, professional, and ethical dimensions of community-based healthcare work. As a result, many systems risk being treated as isolated prototypes rather than as contributions to cumulative design theory.

This study seeks to move beyond system-specific evaluation toward synthesis and theory building. Drawing on longitudinal evidence from iterative deployments, we ask:

\textbf{RQ:} What design guidelines support sustained engagement, learning transfer, and real-world applicability in game-based refresher training systems for Community Health Workers in low-resource healthcare contexts?

To answer this question, we conducted a mixed-methods synthesis of qualitative and quantitative data collected across multiple design cycles with CHWs in India. Through thematic analysis \cite{BraunAndClarke2006} and cross-study comparison, we derived a set of empirically grounded design guidelines that articulate how game-based training systems can be aligned with situated work practices, professional identity, and ethical considerations.

Our findings show that effective game-based training for CHWs depends not only on engaging mechanics but also on contextual realism, hybrid physical–digital interaction, cooperative social structures, explainability, and transparency in data use. These guidelines highlight that game-based interventions must be understood as socio-technical systems embedded in professional and cultural ecosystems rather than as standalone educational tools.

By synthesizing insights across multiple systems and deployments, this paper contributes a conceptual design framework for Interactive Health research. It extends the existing literature on serious games and health training by focusing on the importance of the contextual and ethical dimensions of frontline healthcare work and by offering reusable guidance for future interventions targeting CHWs and similar populations in low-resource settings.

\section{Method}

The study was conducted in rural and semi-urban regions of India, where community health service delivery is primarily facilitated by two female cadres of CHWs: Accredited Social Health Activists (ASHAs), constituted under the National Rural Health Mission, and Anganwadi Workers (AWWs), constituted under the Integrated Child Development Services (ICDS). AWWs operate Anganwadi centers that provide maternal and child care services. Despite distinct institutional affiliations, there is substantial functional overlap between the two cadres, particularly in maternal and child health counseling, growth monitoring, vaccination follow-up, and referral facilitation. These settings are characterized by limited technological infrastructure, intermittent connectivity, and strong socio-cultural influences on health practices. Participants varied in age, education level, and digital literacy, reflecting the heterogeneity of the CHW workforce. Ethical approval was obtained prior to each deployment. Participation was voluntary and did not influence formal performance evaluation or employment status.

This study adopts a Design-Based Research (DBR) \cite{WangHannafinDBR2005} and mixed-methods synthesis approach to derive empirically grounded design guidelines for game-based refresher training systems for CHWs in low-resource healthcare contexts. Rather than evaluating a single intervention, we synthesize evidence from multiple iterative deployments of game-based training systems developed and studied between 2021 and 2024 with ASHAs and AWWs in India. Our methodological goal was to generate generalizable design knowledge by examining recurring patterns across systems, contexts, and time.

Design-Based Research is well-suited to studying complex socio-technical interventions embedded in real-world settings because it emphasizes iterative cycles of design, deployment, evaluation, and theoretical abstraction. Over four years, researchers conducted four major design cycles exploring quiz-based mobile training applications, physical and augmented reality playful activities, digital–physical hybrid card-based games, and location-based games for malnutrition mapping and performance awareness. Each cycle built on insights from the previous one through structured reflection on empirical findings and design breakdowns. This longitudinal structure enabled us to move beyond system-specific outcomes toward the synthesis of design framework.

Participants were recruited through local health departments and partner organizations. Inclusion criteria required participants to be active ASHAs or AWWs with at least one year of work experience and basic familiarity with smartphone use. Across the four design cycles, (n=80) CHWs participated. Some participants (n=45) engaged in multiple cycles, allowing us to observe shifts in perceptions and practices over time. All participants were women.

Data collection combined qualitative and quantitative methods to capture experiential, social, and behavioral aspects of system use. Semi-structured interviews were conducted with (n=42) participants across all design cycles. Interviews explored perceptions of usefulness, engagement, cultural appropriateness, professional legitimacy, and concerns regarding accountability and authority. Interviews were conducted in local languages, audio-recorded, and transcribed for analysis. Field observations were conducted during gameplay sessions and training workshops, with researchers documenting interactional dynamics, moments of confusion or breakdown, peer learning behaviors, and emotional responses such as pride, anxiety, and resistance.

System usage logs were also collected for all digital interventions. These logs included frequency of use, completion rates, error patterns, and response times. For the location-based system, anonymized geographic interaction data and task completion events were recorded. Throughout the study, we documented design rationales, iteration decisions, unexpected outcomes, and failure cases. These supported theory building by linking design intentions with observed field practices.

Data analysis followed an inductive thematic analysis process \cite{BraunAndClarke2006, BraunAndClarke2019} informed by grounded theory guidelines \cite{Charmaz2014}. Analysis began with repeated familiarization with interview transcripts, observation notes, and design notes. Two researchers independently conducted open coding on an initial subset of the data to identify themes related to engagement, learning transfer, social interaction, trust, professional identity, and ethical concerns. Codes were compared and consolidated through discussion, resulting in a shared coding scheme that was subsequently applied to the full dataset using qualitative analysis software NVivo \cite{Jones2014nvivo}. 

Axial coding was conducted on the themes related codes into higher-level conceptual categories such as engagement mechanisms, contextual realism, authority and legitimacy, peer dynamics, and surveillance concerns. We performed a cross-study synthesis by comparing themes across the four design cycles to identify recurring patterns and persistent tensions. Quantitative usage data were used to triangulate qualitative findings by revealing patterns of sustained engagement, drop-off, and interaction frequency. Negative case analysis was explicitly undertaken to examine instances in which game mechanics failed or elicited unintended anxiety or resistance.

Design guidelines were derived through an abstraction process that moved from concrete observations to generalized guidance. The recurring problem–solution patterns were identified across datasets. We then articulated the underlying rationale for these patterns and generalized them beyond specific implementations to formulate transferable design guidelines applicable to interactive health interventions more broadly. Each guideline was supported by evidence from multiple data sources and at least two design cycles.

To ensure rigor and validity, we employed researcher triangulation through collaborative coding and interpretation, data triangulation across interviews, observations, logs, and notes, and member checking with a subset of participants and supervisors to confirm our interpretation. Reflexivity was maintained throughout the study by documenting the researcher's positionality and acknowledging our dual role as designers and evaluators of the systems studied. This approach aligns with calls for meaningful community engagement over superficial data extraction in low-resource contexts \cite{Jones2024slummingit}. This process strengthened the credibility and transferability of the findings.

\section{Results}

Our analysis resulted in eight design guidelines that describe how game-based training systems can support sustained engagement, learning transfer, and contextual appropriateness for CHWs in low-resource settings. These guidelines reflect conflicts between playfulness and professionalism, automation and explainability, and motivation and ethics.

\begin{enumerate}
    \item The first guideline focuses on the importance of contextual realism over abstract correctness. Participants consistently reported that training content felt valuable when it resembled real counseling situations involving families, cultural beliefs, and resource constraints rather than decontextualized quiz questions. Scenario-based narratives enabled CHWs to rehearse responses to conflicts, misinformation, and emotional situations, supporting transfer from game play to field practice.

    \item The second guideline focuses on progressive complexity supported by adaptive feedback. Systems that began with simple tasks and gradually introduced conflicts and challenges were perceived as less intimidating and more motivating. Adaptive feedback allowed participants to learn from mistakes without fear of failure, reinforcing confidence and long-term retention. In contrast, static difficulty levels led to disengagement among both novice and experienced CHWs.

    \item The third guideline focuses on the role of hybrid physical–digital interaction in building trust and social learning. Physical artifacts, such as cards and boards, made digital systems more legible and socially acceptable during group training sessions. These hybrid interactions supported discussion, peer explanation, and collaborative problem solving, transforming individual gameplay into a shared learning activity.

    \item The fourth guideline focuses on the value of location-aware play in improving relevance and accountability. By tying game activities to real geographic areas and household visits, CHWs perceived training as directly connected to their everyday work rather than as an abstract educational exercise. Location-based mechanics supported reflection on coverage gaps and service distribution, strengthening the sense of purpose and applicability.

    \item The fifth guideline focuses on social motivation without surveillance (i.e., systems perceived as punitive for incorrect actions). While participants appreciated opportunities to compare progress with peers, competitive leaderboards and ranking systems often generated anxiety and fear of judgment. Cooperative goals and collective improvement framing were better received and aligned with existing community-based work cultures.

    \item The sixth guideline focuses on explainability over automation. Participants expressed distrust toward systems that only provided correct or incorrect feedback without explanation. Reflection prompts and rationales for answers enabled deeper understanding and reinforced professional competence. Explainability was especially important in scenarios involving counseling and referral decisions.

    \item The seventh guideline focuses on preserving professional identity within playful systems. Excessive gamification, cartoon aesthetics, or rewards risked undermining the seriousness of healthcare work. Designs that balanced playfulness with medical legitimacy and respectful tone were more readily accepted and sustained over time.

    \item The eighth guideline focuses on ethical sensitivity in data collection and performance tracking. Systems that visualized location data or task completion raised concerns about surveillance and blame. Transparency about data use, voluntary participation, and clear separation from formal evaluation processes were essential to maintaining trust.
    
\end{enumerate}

\section{Discussion}

This study contributes a set of empirically grounded design guidelines for game-based refresher training of CHWs in low-resource contexts. These guidelines extend prior work on serious games and interactive health by foregrounding contextual realism, professional identity, and ethical sensitivity as central design considerations. In this section, we situate our findings within existing theoretical and empirical literature and discuss their implications for interactive health research and practice.

\subsection{Situated Learning and Contextual Realism} 

Our findings strongly support theories of situated learning, which argue that knowledge acquisition is inseparable from the social and cultural contexts in which it is applied \cite{LaveWenger1991}. Participants consistently reported greater value in scenario-based and narrative-driven game designs that reflected real counseling encounters rather than abstract biomedical facts. This aligns with prior research demonstrating that experiential and scenario-based simulations improve transfer of learning in health training environments \cite{Glavin2009, Graafland2014}.

Unlike clinical simulation studies conducted in controlled laboratory settings, our work demonstrates how contextual realism must also incorporate cultural beliefs, family dynamics, and resource constraints. This expands the scope of situated learning theory within Interactive Health by emphasizing that realism is not only technical fidelity but also socio-cultural fidelity. These findings suggest that game-based training for CHWs must represent the ambiguities and negotiations that are prevalent in community-based healthcare, rather than idealized clinical procedures.

\subsection{Motivation, Engagement, and Self-Determination Theory} 

Autonomy, competence, and relatedness described in Self-Determination Theory (SDT), relate to motivational mechanisms \cite{DeciRyan1985}. Progressive complexity and adaptive feedback supported participants’ sense of competence by allowing them to learn without fear of failure. Hybrid physical–digital interaction and cooperative play fostered relatedness by encouraging peer discussion and shared problem-solving.

Prior studies of gamification in health training often emphasize points, badges, and leaderboards as motivational drivers \cite{Hamari2014DoesGamification}. Our findings complicate this view by showing that competitive metrics can generate anxiety and resistance in professional health contexts. CHWs perceived ranking systems as forms of surveillance rather than motivation. This aligns with the critiques in HCI4D that warn against importing game mechanics without accounting for local power relations and institutional hierarchies \cite{Sambasivan2010}. Our results suggest that cooperative and collective framing of progress is more compatible with the community-oriented nature of CHWs' work.

\subsection{Professional Identity and Legitimacy in Playful Systems}

Across the previous studies, we observed a conflict to balance between playfulness and professional legitimacy. While participants valued interactive and enjoyable training, excessive gamification risked undermining the seriousness of healthcare work. This aligns with research on professional identity in workplace technologies, which shows that tools must reinforce rather than trivialize occupational roles \cite{Dourish2001}.

In contrast to entertainment-focused games, serious games on interactive healthcare training interventions for CHWs must communicate medical authority and trustworthiness. This requirement distinguishes CHW training games from educational games designed for students or the general population. Our findings highlight professional identity as a design dimension alongside usability and engagement.

\subsection{Explainability and Trust in Decision Support}

Participants’ demand for explanations rather than binary correctness. This aligns with emerging work on explainable AI and transparent decision support systems in healthcare \cite{Doshi2017}. Even in non-AI game-based systems, CHWs sought to understand why an answer was correct to justify their decisions to families and supervisors. Explainability supported reflective learning and reinforced their confidence.

This finding suggests that game-based training systems should be designed as reflective learning environments rather than automated assessment tools. It also resonates with prior Interactive Health studies, which emphasize that trust in health technologies depends on interpretability and alignment with users’ reasoning processes \cite{Amershi2019}.

\subsection{Ethics, Surveillance, and Power}

Location-based and performance-tracking systems introduced ethical tensions related to surveillance and accountability. Participants expressed concern that data collected through games could be used for monitoring or punishment. These concerns align with critiques in digital health and ICTD literature regarding datafication and power asymmetries in low-resource contexts \cite{Barocas2018}.

Our findings demonstrate that ethical considerations cannot be treated as secondary to functionality. Transparency, voluntary participation, and separation from formal evaluation mechanisms were essential for maintaining trust among CHWs. This reinforces the need for value-sensitive and participatory approaches to system design \cite{Friedman2009}.

\subsection{Design Framework}

\begin{figure}[!htp]
    \centering
    \includegraphics[width=1\linewidth]{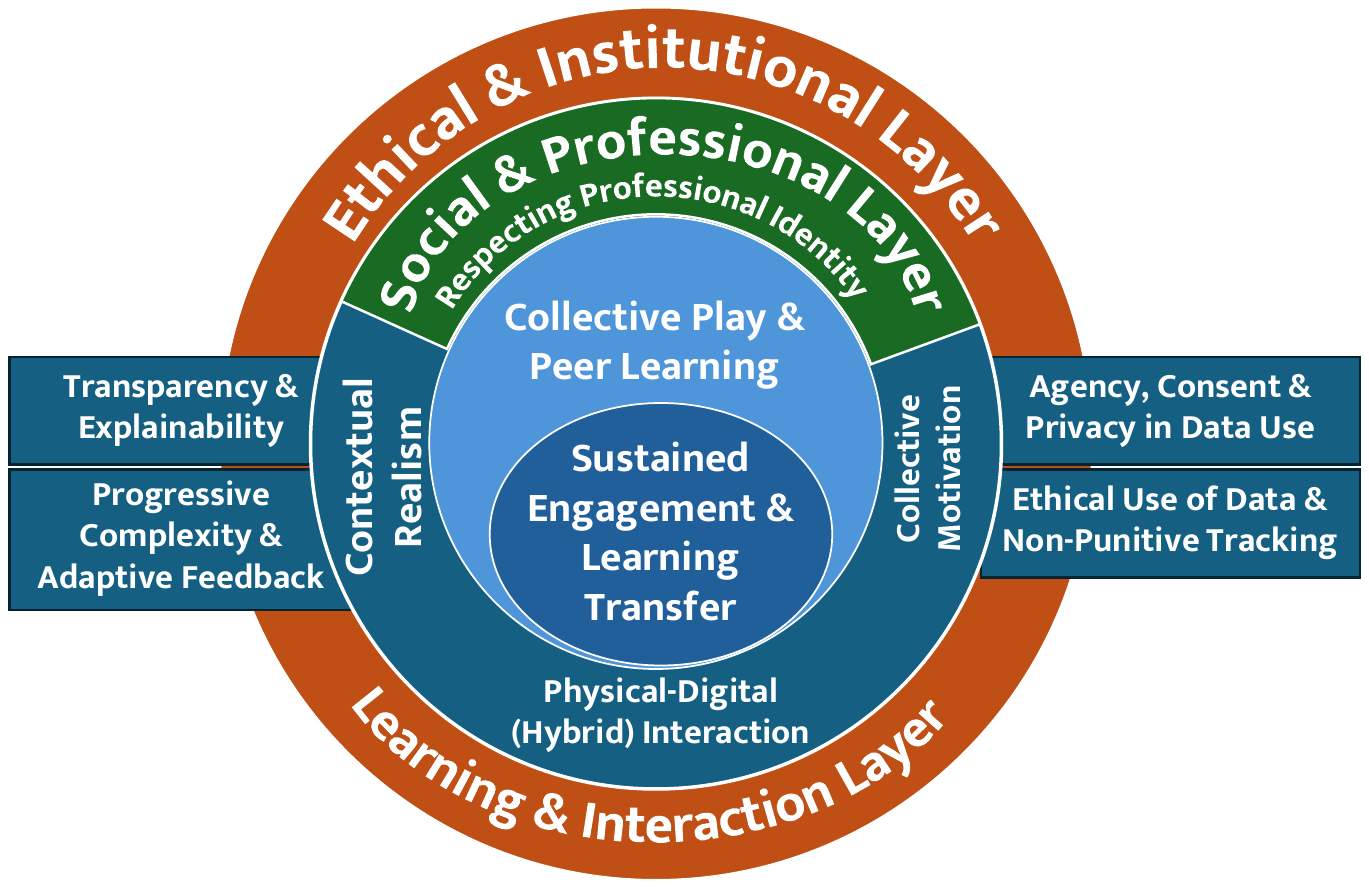}
    \caption{Conceptual Framework for Game-Based CHW Training}
    \label{fig:placeholder}
    \Description[Conceptual Framework for Game-Based CHW Training]{Conceptual Framework for Game-Based CHW Training}
\end{figure}

Based on these design guidelines, we propose a conceptual framework for game-based refresher training in low-resource healthcare contexts. Figure \ref{fig:placeholder} illustrates a three-layer conceptual design framework for game-based refresher training of Community Health Workers.

At the learning layer, contextual realism, progressive complexity, and explainability support knowledge acquisition and transfer to real-world counseling practices. These elements ensure that training content aligns with situated work rather than abstract curricula. At the interaction layer, hybrid physical–digital interaction, location-aware play, and cooperative social mechanics shape how CHWs engage with the system and with one another. This layer emphasizes that learning is socially constructed and embedded in collective practice. At the ethical and professional layer, respect for professional identity and sensitivity to surveillance and data use govern acceptance and sustainability. This layer addresses trust, legitimacy, and power relations that shape long-term adoption.

The framework illustrates how effective game-based training emerges from the alignment of these three layers rather than from isolated game mechanics. Design breakdowns occur when one layer dominates at the expense of others, such as when engagement features undermine professional legitimacy or when monitoring features reduce trust.

\section{Limitations and Future Work}

This study has several limitations that should be considered when interpreting the findings. 

The research was conducted primarily with ASHAs and AWWs in specific regions of India. While these populations represent a large and critical segment of CHWs, the socio-cultural and institutional conditions in which they operate may differ from those in other countries or healthcare systems. As such, the design guidelines derived here should be understood as analytically generalizable rather than universally applicable. Further studies are needed to examine how these guidelines translate to other contexts, such as urban community health programs or refugee health initiatives.

Although the study draws on data collected over multiple years, each deployment was limited in duration and scale. Constraints related to funding, institutional access, and public health disruptions influenced the continuity of some field studies. While this work offers longitudinal insight at the level of design iteration, it does not capture multi-year learning outcomes or direct measures of patient health impact. Future research should investigate how sustained use of game-based training systems influences long-term knowledge retention, behavioral change, and service quality.

The data used in this synthesis relies on self-reported experiences, observations, and qualitative feedback from participants. While these methods are well-suited for uncovering design insights and contextual factors, they may be subject to social desirability bias and researcher interpretation. Complementary methods, such as controlled comparisons with conventional training programs or objective performance metrics, would strengthen the empirical grounding of future studies.

The tools presented in this study were developed and evaluated within a research-driven design process rather than institutionalized government training programs. This limits conclusions about scalability, maintenance, and integration with existing digital health infrastructures. Future work should explore partnerships with health authorities to examine how game-based training tools can be embedded into official training curricula and supported over time.

Ethical considerations surrounding data collection, monitoring, and performance visualization remain an open area for further inquiry. While our design guidelines emphasize transparency and agency, additional research is needed to understand the long-term implications of location tracking, performance metrics, and gamified evaluation on worker autonomy and trust. Participatory governance models and policy-level interventions may be necessary to ensure that interactive training systems empower rather than discipline frontline workers.

Future research directions include expanding the design framework to incorporate adaptive personalization, exploring AI-supported feedback mechanisms that remain interpretable to users, and investigating how hybrid physical–digital systems can be deployed at scale with minimal technical support. Longitudinal studies examining the transfer of learning into everyday work practices and patient outcomes will be essential to establishing the true impact of game-based refresher training in community healthcare.

\section{Conclusion}
This study presented a longitudinal synthesis of game-based refresher training systems designed for Community Health Workers (CHWs) in India, drawing on iterative design, deployment, and evaluation. Through the analysis of multiple digital refresher tools we derived empirically grounded design guidelines that articulate how interactive health interventions can support sustained engagement, learning transfer, and contextual relevance in low-resource healthcare settings.

Our findings demonstrate that successful game-based training for CHWs extends beyond the implementation of engaging game mechanics. Instead, effectiveness emerges from alignment with situated work practices, professional identity, and social learning structures. Systems that incorporated contextual realism, cooperative interaction, and embodied physical components were more readily adopted and perceived as legitimate training tools. Equally important was the role of explainability and transparency in supporting trust, particularly in systems that collected or visualized performance and location-based data. These insights highlight the need to conceptualize serious games for health training as socio-technical interventions within professional and cultural ecosystems rather than as isolated educational tools.

By synthesizing evidence across multiple deployments, this study contributes design guidelines and conceptual design framework to the Interactive Health and HCI communities. The design guidelines provide reusable guidance for researchers and practitioners seeking to design training systems for CHWs in resource-constrained environments. Beyond the specific context of India, these guidelines are relevant to similar community-based healthcare programs in other low and middle income countries (LMICs), where training infrastructure faces comparable challenges of scale, diversity, and sustainability \cite{Borda2023}.

This study advances theoretical understanding of how playful and game-based systems can support professional learning in non-traditional settings. It extends existing theories of situated learning and motivation by demonstrating how play can coexist with professional seriousness when grounded in real-world tasks and ethical considerations. The work also contributes to value-sensitive design in Interactive Health by focusing on concerns of dignity, agency, and transparency in the design of data-driven training systems for vulnerable worker populations.


\begin{acks}
We thank the Community Health Workers and their supervisors for their participation and support in this study. We also acknowledge the contributions of the partner NGOs, their teams, and associated organizations. We are grateful to the reviewers for their constructive feedback. This work was supported by the Science and Engineering Research Board (SERB), the Federation of Indian Chambers of Commerce and Industry (FICCI), and UNICEF, New Delhi.
\end{acks}

\bibliographystyle{ACM-Reference-Format}
\bibliography{sample-base}

\end{document}